\documentclass{aa}  

\usepackage{graphicx}
\usepackage{txfonts}
\def\ch{\textit{Chandra}}
\def\gaia{\textit{Gaia}}

\newcommand{\furl}[1]{\footnote{\url{#1}}}

\begin{document}

   \title{Using radial velocities to reveal black holes in binaries: a test case}

   %\subtitle{}

   \author{M. Clavel\inst{\ref{inst1}} \and G. Dubus\inst{\ref{inst1}} \and J. Casares\inst{\ref{inst2},\ref{inst3}} \and C. Babusiaux\inst{\ref{inst1}}
          }

   \institute{Univ. Grenoble Alpes, CNRS, IPAG, F-38000 Grenoble, France\label{inst1}\\
              \email{maica.clavel@univ-grenoble-alpes.fr}
    \and
     Instituto de Astrof\'isica de Canarias, E-38205 La Laguna, Tenerife, Spain\label{inst2}
     \and
     Departamento de Astrof\'isica, Universidad de La Laguna, E-38206 La Laguna, Tenerife, Spain\label{inst3}
             }

   \date{Received Sept 1, 2020; accepted November 1, 2020}%Received September 15, 1996; accepted March 16, 1997}

  \abstract
   {} %leave it empty if necessary  
   {Large radial velocity variations in the LAMOST spectra of giant stars have been used to infer the presence of unseen companions. Some of them have been proposed as possible black hole candidates. We test this selection by investigating the classification of the one candidate having a known X-ray counterpart (UCAC4 721-037069).}
   {We obtained time-resolved spectra from the Liverpool Telescope and a 5\,ks observation from the \textit{Chandra} observatory to fully constrain the orbital parameters and the X-ray emission of this system.} 
   {We find the source to be an eclipsing stellar binary that can be classified as a RS CVn. The giant star fills its Roche Lobe and the binary mass ratio is greater than one. The system may be an example of stable mass transfer from an intermediate-mass star with a convective envelope.}
  {Using only radial velocity to identify black hole candidates can lead to many false positives. The presence of an optical orbital modulation, such as observed for all LAMOST candidates, will in most cases indicate that this is a stellar binary.}


   \maketitle
%
%-------------------------------------------------------------------

\section{Introduction}

Known compact binaries with neutron stars (NS) or black hole (BH) as compact objects are currently identified thanks to their bright X-ray outbursts  \citep{mcclintock2006,corralsantana2016} or from follow-up observations of X-ray surveys \citep[e.g.][]{walter2015}. However, these bright X-ray sources most likely represent only a small fraction of the overall population of compact binaries, with a larger number of X-ray faint objects missed by the current approach because of long quiescent periods in between outbursts \citep{dubus2001, yungelson2006, yan2015} or because they are persistently very faint \citep{menou1999, king2006}.

Radial velocity variations in optical spectra could be a mean to uncover the population of quiescent compact object having a stellar companion \citep{casares2014, giesers2018, thompson2019, makarov2019}. 
Recently, \citet{gu2019} selected spectra of bright late-type giants obtained at different times by the LAMOST spectral survey and identified seven candidates with large radial velocity variations, compatible with the presence of an unseen solar mass companion of $>3\rm\,M_\odot$. One of them (UCAC4 721-037069, which corresponds to their source~\#4, hereafter GS4) has a possible X-ray counterpart and is presented by \citet{gu2019} as their best black hole candidate.

In this paper, we investigate the data available, including two dedicated observation programs, to identify the nature of this intriguing source. The optical spectra, the orbital parameters and the properties of the X-ray counterpart of GS4 are described in Section~\ref{sec:data}. The identification of this source as a RS CVn is then discussed in Section~\ref{sec:discussion}, along with considerations on how constraints on the orbital period could further help to discriminate between this type of sources and quiescent black holes having a stellar companion.

\section{Optical and X-ray constraints}
\label{sec:data}

GS4 corresponds to Gaia DR2 273187064220377600. Located at ${\rm R.A.\  (J2000)} = 04^h 56^m 12.7739^s$ and ${\rm Dec.\ (J2000)} = 54^\circ 00' 21.2768''$, this source has a parallax of $1.07\pm0.03$\,mas, which corresponds to a distance of $0.91\pm0.03$\,kpc \citep{bailerjones2018}.

The All-Sky Automated Survey for Supernovae \citep[ASAS-SN,][]{shappee2014,kochanek2017,jayasinghe2019} and \textit{Gaia} DR2 \citep[][]{gaiacollaboration2016,gaiacollaboration2018} provide aperture photometry light curves of GS4. As previously reported by \citet{zheng2019}, these light curves have a strong modulation with a period $p \approx 5.2\rm\,days$, and the morphology of GS4's folded light curve is typical of eclipsing binaries  (see Fig.~\ref{fig:lc+rv}, top, and Sect.~\ref{sec:mcmc}). 

The extinction towards GS4 can be estimated from the 3D map of interstellar dust reddening derived from Pan-STARRS 1 and 2MASS surveys \citep{green2018}. At 910\,pc the reddening is $E(g-r) = 0.35\pm0.02$, which corresponds to an extinction $A_v = 1.06\pm0.06$.\footnote{Using \textit{Gaia} and 2MASS surveys, \citet{lallement2019} derived an extinction of 1.4 mag in this direction and at this distance. Using this value would require a higher luminosity from the binary components to account for the observed flux: it would increase the tension between spectral type and luminosity, furthering the need for additional light (see Sect.~\ref{sec:mcmc} and \ref{sec:nature}).} This value converts into a column density $N_{\rm H} \approx 2.3\times10^{21} \rm \,cm^{-2}$ \citep[see][]{guver2009}.

\subsection{Time resolve spectroscopy and radial velocity}
\label{sec:RV}
In agreement with the star's position in the \textit{Gaia} Hertzsprung–Russell diagram, the LAMOST spectra of GS4 revealed a late type giant having radial velocity variations as large as $\Delta V_r = 127.2 \pm 7.8\rm\,km\,s^{-1}$ \citep{gu2019,zheng2019}. 
We obtained refined radial velocity measurements using the medium resolution spectrograph FRODOSpec at the robotic Liverpool Telescope \citep[LT,][]{steele2004, barnsley2012} to establish the orbital parameters of this source.

The spectrograph is fed by a fiber bundle array consisting of $12\times12$ lenslets of 0.82$\arcsec$ each which is reformatted as a slit. FRODOSpec was operated in high resolution mode, providing a spectral resolving power of $R\sim5500$ in the blue arm and $R\sim5300$ in the red arm. The spectral coverage was 3900--5215\,\AA ~and 5900--8000\,\AA, respectively. A total of twenty six 1200\,s spectra were obtained in each arm, with a cadence of 1-3 spectra per night between 2019 Aug 28 and 2019 Dec 1. 
For the sake of radial velocity analysis we also obtained a 300\,s spectrum of the K0III standard HD 210185 on the night of 2019 Sep 17, using the same spectral configuration. The FRODOSpec pipeline produces fully extracted and wavelength calibrated spectra with rms$\le$0.1\AA\ above 4400\,\AA. The analysis presented in this paper has been performed with the FRODOSpec pipeline products.

Radial velocities were extracted through cross-correlation of every GS4 red arm spectrum with the K0III template  in the spectral range 6300-7900\,\AA, after masking the main telluric and IS absorption bands. The H$_{\alpha}$ line was also excluded as some spectra show evidence for variable narrow-emission, reminiscent of chromospheric activity.

The spectrum obtained for GS4 is compatible with a K0III star having a temperature  of the order of 4800\,K  \citep[consistent with the effective temperature derived from LAMOST DR6, see e.g.][]{gu2019} and a periodogram analysis of the radial velocity points shows a clear peak at  5.2081(34)\,d, in excellent agreement with the ASAS-SN photometric period (see below).
The radial velocity curve is shown in Figure~\ref{fig:lc+rv} (Bottom). Modeling the light curve with a simple sinusoid gives the systemic velocity $\gamma = -7.8\pm1.1\rm\,km\,s^{-1}$  
and the semi-amplitude radial velocity $K_1 = 69.0 \pm 1.7\rm\,km\,s^{-1}$. Together with the orbital period of the system determined by phase dispersion minimization on the ASAS-SN V-band light curve, $P_{\rm orb}=5.20906\pm0.0005\,\mathrm{d}$, this corresponds to a binary mass function $f=0.18\pm0.01\rm\,M_{\sun}$.

We observe a shift of 0.04 in phase between the ASAS-SN light curve and the LT radial velocity curve (see Fig.~\ref{fig:lc+rv}, bottom). This corresponds to five hours, so we can rule out any instrumental effect\footnote{Both LT and ASAS-SN curves are provided in HJD and the ASAS-SN photometry aligns well with publicly available data from other observatories (e.g.\ \textit{Gaia}).}. A persistent asymmetry in the stellar radiation field (due to e.g.\ star spots, see also Sect.~\ref{sec:nature}) could be a possible physical origin, but we have no evidence for it. We note that an asymmetric flux distribution combined with eclipses has also been discussed as the possible origin of a similar phase lag detected in the black hole X-ray binary XTE\,J1118$+$480 during its decay to quiescence \citep{mcclintock2001}.

\begin{figure}
	\includegraphics[width=\columnwidth]{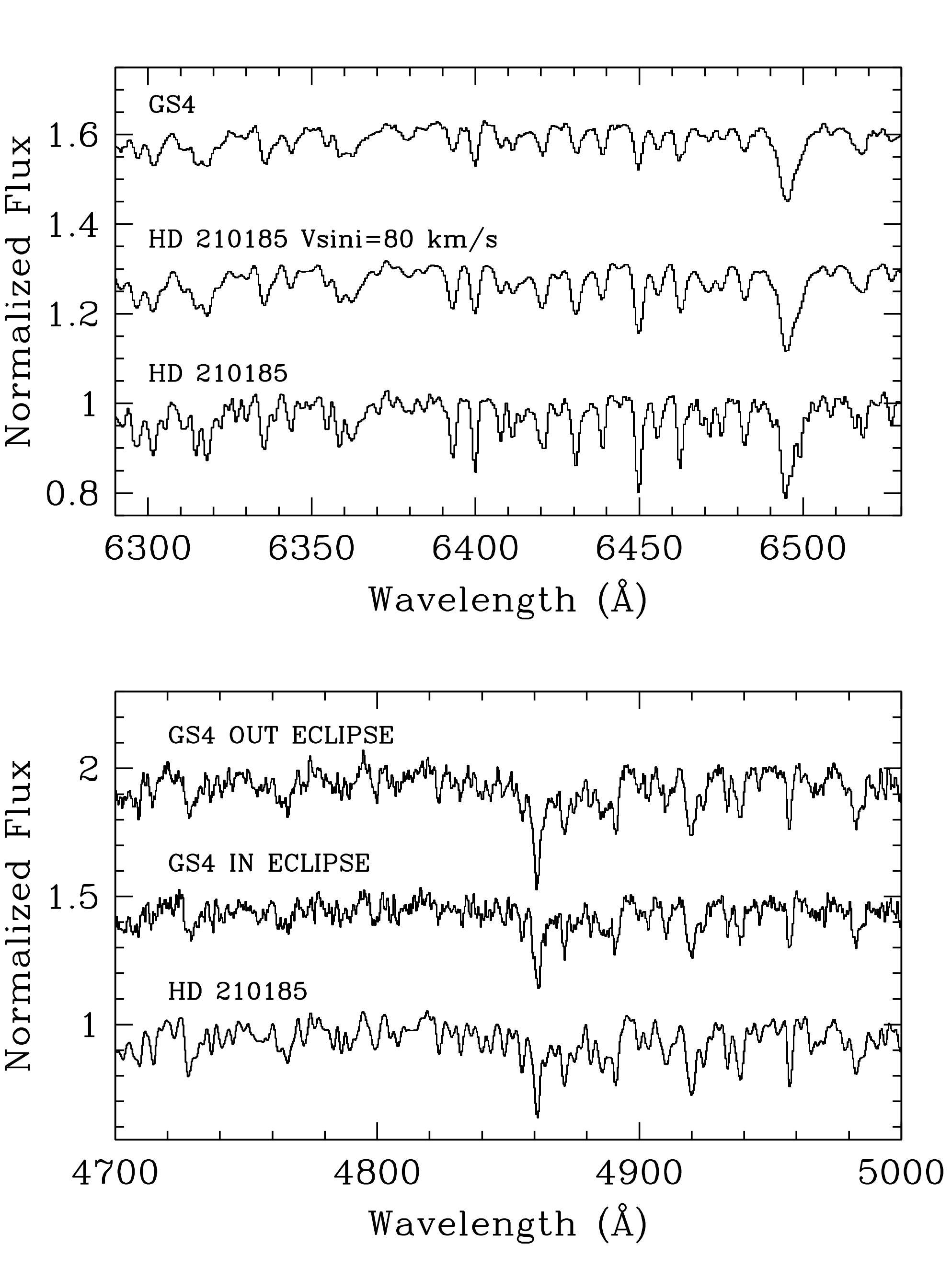}
	\caption{\textit{(Top)}  From bottom to top, the red-arm LT spectrum of HD 210185 (the K0III template), the same spectrum broadened by 80 km/s and the doppler-corrected average spectrum of GS4. \textit{(Bottom)} From bottom to top, the blue-arm LT spectrum of the K0III template in comparison with the doppler-corrected averages of GS4 obtained during the primary eclipse (photometric phase $0.00\pm0.05$) and outside the eclipse (rest of phases). The top spectrum shows the H$_\beta$ line (at $4861\,\AA$) to be slightly stronger (relative to the nearby metallic lines), possibly because of the contribution from the A-type secondary. This one is eclipsed in the middle spectrum, which seems to show a better match to the K0III spectrum. Scaling factors have been applied to the three spectra so that the equivalent width of the metallic lines in the 4900-5000\,$\AA$ range is the same.} 
	\label{fig:spec}
\end{figure}

In order to constrain the rotational broadening of the K0III star we have broadened the red-arm spectrum of the template star (HD 210185) from 0 to 120  km s$^{-1}$ in steps of $2\rm\,km\,s^{-1}$, using a Gray profile \citep{gray1992} with a linear limb-darkening law with coefficient $\epsilon=0.75$, appropriate for the wavelength range and spectral type of our star \citep[see][]{alnaimiy1978}. The broadened versions of the template star were multiplied by fractions $f<1$, to account for the fractional contribution to the total light, and subsequently subtracted from the Doppler corrected average of the the 26 red spectra of GS4, using our orbital solution. A $\chi^{2}$ test on the residuals yields $V \sin i = 80 \pm 1\rm \, km\,s^{-1}$. The comparison of GS4 spectrum with the best broadened version of the template star is shown in Figure~\ref{fig:spec} (top).

\begin{figure}
    \includegraphics[width=\columnwidth]{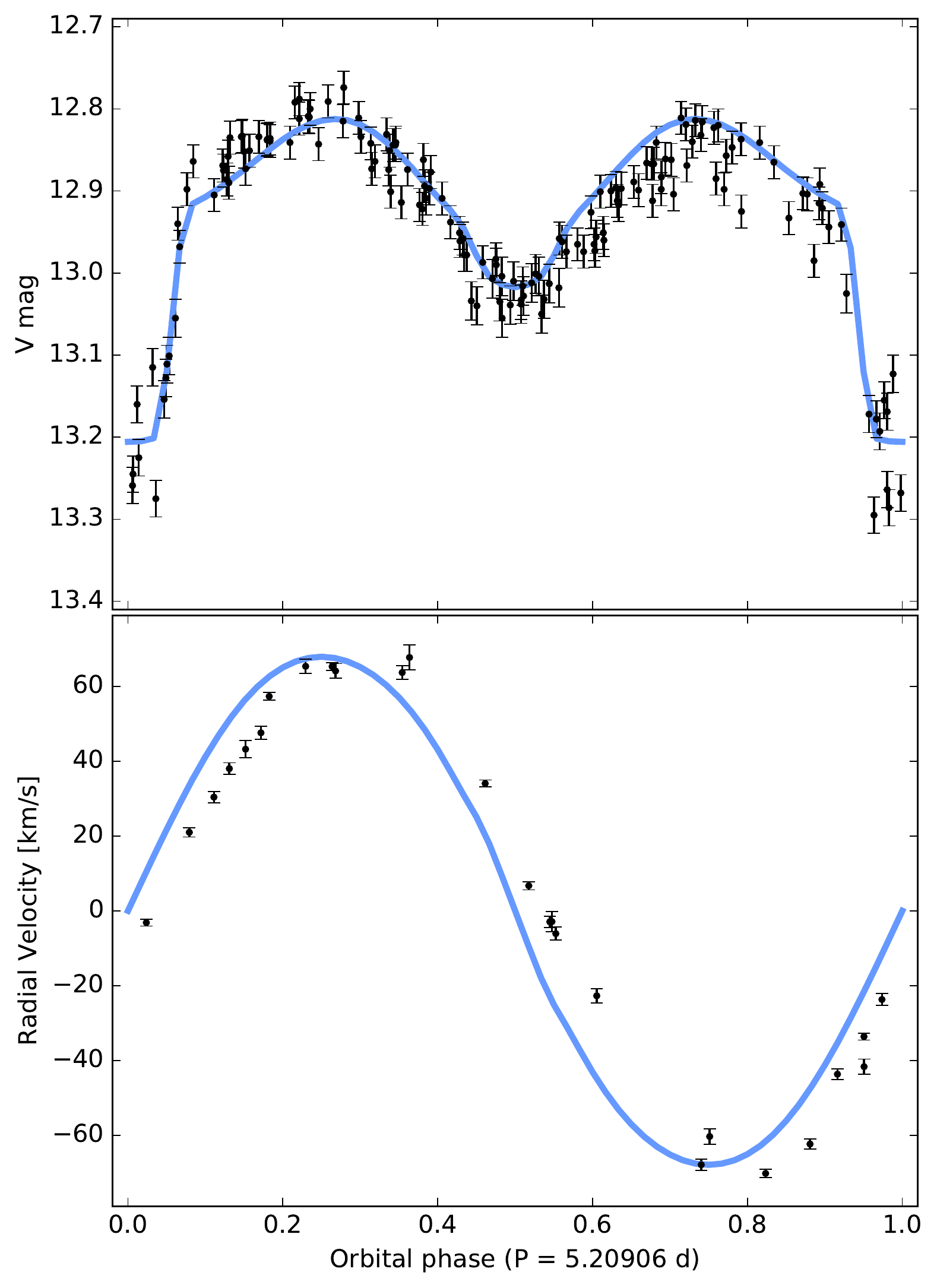}
    \caption{\textit{(Top)} ASAS-SN light curve of GS4, folded using the period $p=5.20906\pm0.0005\rm\,d$ and $t_0 = 2457384.6\rm\,HJD$. \textit{(Bottom)} Corresponding radial velocity of the system as observed by LT. We do see a 0.04 phase shift between the radial velocity curve and the ASAS-SN light curve but we have no clear explanation for it (see Sect.~\ref{sec:RV}). 
    The blue curves correspond to the best fit model obtained with PHOEBE on the ASAS-SN light curve when fixing the semi-amplitude radial velocity of the system to the one measured from the LT curve (see Sect.~\ref{sec:mcmc}).} 
    \label{fig:lc+rv}
\end{figure}

\subsection{System parameters derived from the lightcurve}
\label{sec:mcmc}

We model the ASAS-SN light curve using the eclipsing binary modeling software PHOEBE\footnote{\url{http://phoebe-project.org/}} with PHOENIX model atmosphere \citep{jones2020}, in combination with the MCMC sampler {\em emcee} \citep{foremanmackey2013}. We fixed the orbital period to the value derived from phase dispersion minimization (see Section~\ref{sec:RV}).  
We set the reddening to $A_{V}=1.06$, derived from the map of \citet[][see above]{green2018}. The free parameters of the fit were the system inclination $\cos i$, the primary (K0III) star mass $M_{1}$, the effective temperatures of the binary components $T_{1,2}$, the ratio of primary star radius $R_{1}$ to Roche lobe radius $R_{L}$, the radius of the secondary (unseen) star $R_{2}$, the parallax $\pi$. The mass of the secondary star $M_{2}$ was set from $M_{1}$, $i$ and $K_{1}=69\,\mathrm{km\,s}^{-1}$. We used flat priors except for the parallax $\pi$ where we used a gaussian prior ($1.0981\pm 0.0421\,\mathrm{mas}$). The parallax takes into account the parallax zero point of 0.03 mas and the error increase suggested by Lindegren et al.\furl{https://www.cosmos.esa.int/web/gaia/dr2-known-issues#AstrometryConsiderations}. The sampler ran for 30,000 steps, after which we found the chain had converged based on the sampled parameters and the autocorrelation time. We discarded the initial 5000 steps (burn-in phase). The errors quoted in Tab.~\ref{tab:sysparam} are based on the 10th and 90th percentiles. A major source of uncertainty is the adopted reddening. We did not attempt to model light curves in different bands. Taking $A_{v}=0$ significantly lowers the effective temperatures ($T_{1}\approx 5060$\,K and $T_{2}\approx 7860\,\mathrm{K}$) and $R_{2}$ to 1.3$\,\mathrm{R}_{\odot}$, lowers slightly $M_{1}$ to 1.6$\,\mathrm{M}_{\odot}$, but has no impact on $i$ and $R_{1}/R_{\rm L}$ since these are constrained by the light curve shape. In all cases, the primary nearly fills its Roche lobe with $R_1 = 7.5\pm0.5\rm\,R_\sun$.

\begin{table}
	\centering
	\caption{Parameters of the binary system GS4 obtained from a fit of the ASAS-SN V-band light curve using PHOEBE. The errors are based on the 10th and 90th percentiles (see Sect.~\ref{sec:mcmc}). Part of these parameters are likely slightly off compared to the true values due to an extra-light contamination (see Section~\ref{sec:nature}).}
	\begin{tabular}{c c c}
		\hline \hline
		Parameters & units & best fit values\\
		\hline
		$i$ & $^\circ$ & $78.5^{+1.2}_{-0.9}$\\
		$M_{\rm1}$ & M$_\sun$ & $1.9\pm0.3 $\\
		$M_{\rm2}$ & M$_\sun$ & $1.2 \pm 0.1$\\
		$R_{\rm1}$/R$_{\rm L}$ &  & $0.982^{+0.005}_{-0.008}$\\
		$R_{\rm2}$ & R$_\sun$ & $1.7 \pm 0.2$\\
		$T_{\rm1}$ & K & $5900^{+200}_{-150}$\\
		$T_{\rm2}$ & K & $8800^{+500}_{-400}$\\
		$\pi$ & mas & $1.14\pm 0.05$\\

		\hline
	\end{tabular}
	\label{tab:sysparam}
\end{table}

\subsection{X-ray counterpart}
\label{sec:xray}
As pointed out by \citet{gu2019}, GS4 is compatible with the position of the \textit{ROSAT} source 2RXS\,J045612.8$+$540024 \citep[$4''$ away from GS4, with a $12''$ X-ray position uncertainty,][]{voges2000} with an association likelihood of 55\% \citep{Flesch2016}. In the 0.1--2.4\,keV range, this X-ray source is rather hard with an X-ray flux of $F_{\rm0.1-2.4keV} = 4.4\times10^{-13}$\,erg\,cm$^{-2}$\,s$^{-1}$. 
To test this association, we obtained 5\,ks of \ch/ACIS-I observation (Obs.~ID 22399) targeting the \textit{Gaia} position of GS4. This X-ray observation spanned from MJD 58728.017 to 58728.075, which corresponds to phase $\approx0.02$ in Fig.~\ref{fig:lc+rv}: it was performed during the primary eclipse.

The data was reduced using CIAO software v.4.12 along with the calibration database CALDB v.4.9.0. \textit{Chandra\_repro} was used to produce the clean event file and then \textit{wavdetect} (default parameters) was  run to detect all sources in the field of view. Sixteen sources are detected, including one source consistent with the \gaia\ position of GS4 detected with a significance of 18$\sigma$. The X-ray source coordinates are ${\rm R.A.} = 04^h 56^m 12.76^s$, ${\rm Dec.} = 54^\circ 00' 21.2''$. The source is therefore named CXO\,J045612.8$+$540021. Its position accuracy is limited by \ch\ absolute astrometric accuracy and is therefore $\sim 1''$. 

We then extracted the source spectrum using \textit{specextract} and a circular region of radius $r=2''$ centered on the position of CXO\,J045612.8$+$540021, while the background spectrum was extracted from a annulus region between $5''$ and $20''$ and centered on the same coordinates. A total of 36 counts are detected from the source in the 0.5--8\,keV energy range (there are 10 counts in the background region in the same energy range, so we expect only 0.1 background count in the source region). The source spectrum was binned to have at least 5 counts per bin, and then background subtracted. We note that the last bin from 5 to 8 keV only contains one count. We therefore use Gehrels weights \citep{gehrels1986} to better account for the error bar in each bin. Spectral fits are performed with \texttt{Xspec} v.12.10.1 and the errorbars correspond to the 90\% confidence interval. The absorption is modeled with \textit{tbabs}, using Wilms abundances \citep{wilms2000}, and fixed to $N_{\rm H} = 2.3\times10^{21}\rm\,cm^{-2}$ (see beginning of Sect.~\ref{sec:data}). 

We first tested a power-law model to fit the source spectrum. The best fit gives a reduced $\chi^2_r = 0.4$ (six degrees of freedom, d.o.f.) and a photon index $\Gamma=1.6^{+1.0}_{-0.8}$. The source spectrum is somewhat better fitted  by a black body model ($\chi^2_r=0.3$, six d.o.f.) with a temperature $kT = 0.8 ^{+0.4} _{-0.3} \rm \,keV$ and the 0.5--10\,keV flux of the source is then $F_{\rm0.5-10keV} = 9.2^{+5.5}_{-4.8} \times 10^{-14}\rm\,erg\,cm^{-2}\,s^{-1}$, which at a distance of 0.91\,kpc translates into a luminosity $L_{\rm0.5-10keV} = (9\pm5) \times 10^{30}\rm\,erg\,s^{-1}$. 

Extrapolating the best fit model to the \textit{ROSAT} energy range, we find a flux that is $F_{\rm 0.1-2.4keV} = 3.4\times10^{-14}\rm\,erg\,cm^{-2}\,s^{-1}$, i.e. more than one order of magnitude below the \textit{ROSAT} detection in 1990.  While we cannot exclude that the \textit{ROSAT} source could be a transient unrelated to GS4 and now under the detection limit of our 2019 \textit{Chandra} observation, it is also likely that the source has varied, either due to the eclipsing pattern and/or a longer term variability.

%--------------------------------------------------------------------

\section{Discussion}
\label{sec:discussion}

\subsection{Nature of the unseen companion}
\label{sec:nature}

The orbital model fitting the orbital light curve of GS4, presented in Section~\ref{sec:mcmc}, describes a stellar binary composed of a giant star nearly filling its Roche lobe and a hotter but smaller companion. At phase 0, the giant star is totally eclipsing its companion, while at phase 0.5 the companion is in front of the giant star and only partly eclipsing it. The deformed shape of the giant star produces the ellipsoidal modulation in addition to the eclipse at phase 0. Therefore, the fit to the orbital modulation constrains very well the inclination and the Roche Lobe filling factor of the primary.  The rotational broadening ($\approx80\rm\,km\,s^{-1}$) is compatible with the primary star being tidally locked and synchronized with $R_1\approx 8\rm\,R_\sun$, as expected since it nearly fills its Roche Lobe. 

Under the assumption that the K0III star fully fills its Roche lobe and is tidally locked in its orbit, the binary mass ratio $q$ can be derived through the expression $V \sin i \simeq 0.462 K_1 q^{1/3} (1+q)^{2/3}$ \citep{wade1988}. In our case $q=M_1/M_2$, where $M_1$ is the mass of the K0III star, and $K_1=69.0\pm1.7$ so we find $q=1.89 \pm 0.07$. This is slightly larger but consistent with the mass ratio derived through the PHOEBE modeling (Table~\ref{tab:sysparam}).

From the orbital evolution of the veiling factor, i.e. the fractional contribution of the K0III template spectrum to the total light, we find that the K0III spectrum contributes 60\% outside the primary eclipse and 91\% at phase 0. 
This means that the secondary accounts for $\approx 30$\% of the total flux and suggests the presence of additional light in the system that is not accounted for in the fit of the orbital modulation. Such additional light could also explain why the temperature $T_1\approx 5800\rm\,K$ derived from the fit is incompatible with the K0III spectral classification ($\approx 4800\rm\,K$). Since the primary star is nearly filling its Roche Lobe, a mass transfer creating an accretion disk around the secondary star is the most likely explanation for this additional source of light.

The mass, temperature and radius of the secondary derived from the orbital fit suggest an A or F-type main sequence companion. 
This identification of the unseen companion is fully consistent with the fact that it is not detected spectroscopically. It is worth noting that the spectrum of the secondary is hinted at phase $\sim0.5$ by the large depth of the Balmer lines (specially H$\beta$ and H$\gamma$ in the blue-arm LT spectra), relative to the nearby metallic lines, as compared to a typical K0 spectrum (see Fig.~\ref{fig:spec}, bottom). 

These optical constraints and the X-ray emission described in Section~\ref{sec:xray}, are fully consistent with a chromospherically active giant star in a semidetached binary and we therefore classify GS4 as a RS CVn \citep[see e.g.][for X-ray properties of such systems]{pandey2012}. These systems do display variability in their X-ray and orbital modulation light curves due to the activity of the giant star. 
We note that the binary mass ratio is quite large ($\sim$1.6--1.9). This value approaches (but does not exceed) the critical mass ratio for unstable mass transfer from intermediate-mass donor stars with convective envelopes, as computed in recent simulations \citep{misra2020}.
In addition, the asymmetry of the light curve seen at phase 0.8--0.9 (Fig.~\ref{fig:lc+rv}, top) is fairly typical for RS CVn and is explained by the presence of starspots at the surface of the giant star.

\subsection{Radial velocities as probe for quiescent black holes}
\label{sec:RVsurvey}

\begin{figure}
\centering
\includegraphics{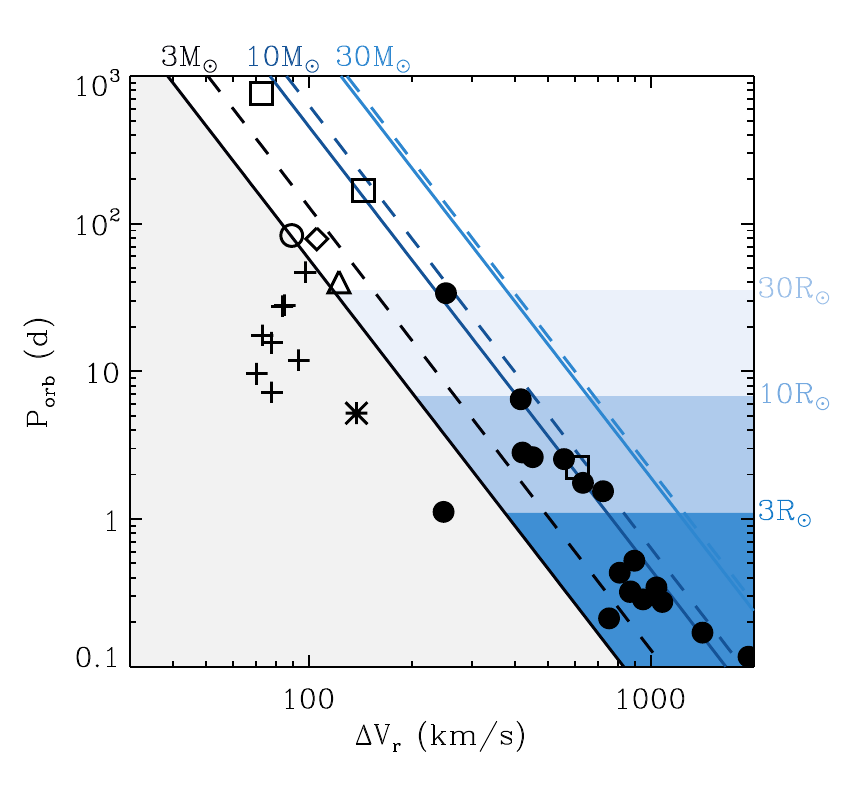} 
\caption{Black hole candidates with solar mass companions in the orbital period -- radial velocity amplitude plane. Full (resp. dashed) lines show the lower limit on $P_{\rm orb}$ according to Eq.~\ref{eq:ll} sufficient to have a 3, 10 or 30\,M$_{\odot}$ black hole in circular orbit with a 3\,M$_{\odot}$ stellar companion (resp. 1\,M$_{\odot}$ companion). Colored areas indicate where a 3, 10, or 30\,R$_{\odot}$ stellar companion would overfill its Roche lobe. Candidate black holes from radial velocity measurements: low-mass X-ray binaries (black dots); APOGEE candidate (open circle); LAMOST LB-1 candidate (open diamond); MUSE candidates in NGC 3201 (open squares); HR 6819 candidate (open triangle) and the latest LAMOST selection of giant stars having large $\Delta V_r$ and optically invisible companions (crosses), including GS4 (star). See text for details.}
\label{fig:limit}
\end{figure}

Population studies indicate that there may be from $10^{3}$ to $10^{5}$ black holes with solar mass companions  (e.g. \citealt{shao2019,olejak2020}). The $\approx 60$ black hole X-ray binaries sample only a fraction of this binary population \citep{corralsantana2016}. 
With a duty cycle $\leq 0.02$ \citep{yan2015} many are hiding in quiescence, the accretion disk building up mass before the next outburst. Many more may be non-accreting if the system was unable to evolve to shorter periods during the common envelope phase or through angular momentum losses \citep{dekool1987, iben1995, podsiadlowski2003, yungelson2006, shao2019}. These binaries can therefore provide important insights into black hole evolution, even though they harbor only a small fraction of the $\sim 10^{8}$ black holes estimated to reside in our Galaxy. 

Radial velocity surveys of field or globular cluster stars can provide dynamical evidence for invisible companions \citep{giesers2018, zheng2019, gu2019, yi2019, wiktorowicz2020}. For a circular binary, 
\begin{equation}
 M_{\rm bh}= \frac{P_{\rm orb} K_{\star}^3 (1+q)^2} {2\pi G\sin^3 i}
\end{equation}
where $K_{\star}$ is the semi-amplitude of the radial velocity modulation of the stellar companion and $q=M_{\star}/M_{\rm bh}$. If a radial velocity difference $\Delta V_r$ is measured between two epochs, a sufficient condition to have a $>3 {\rm M}_{\odot}$ unseen companion is
\begin{equation}
 \frac{P_{\rm orb} \Delta V_r^3 (1+q)^2} {16\pi G}\geq 3  {\rm M}_{\odot}
 \label{eq:ll}
\end{equation}
We plot this limit in Fig.~\ref{fig:limit} for $M_{\rm bh}=3$, 10 and 30\,M$_{\odot}$, assuming $M_{\star}=1$ (dashed lines) or 3 (full lines)\,M$_{\odot}$. It is important to highlight that these lines would be shifted towards the left for systems that are known to have a low inclination. The limit below which the star will overfill its Roche lobe are also highlighted. Overlaid to this plot, we show black hole candidates based on radial velocity measurements: the black hole low-mass X-ray binaries listed in Tab.~A4 of \citet{corralsantana2016}, the black hole candidate found from the APOGEE survey \citep{thompson2019}, the candidate LB-1 found in the LAMOST survey (note the companion is a B star, \citealt{liu2019}, and follow-up discussions by e.g.\ \citealt{irrgang2020,shenar2020}), the black hole candidates found in MUSE observations of the NGC 3201 globular cluster (some have non zero eccentricity, \citealt{giesers2019}) and the non accreting black hole candidate in the triple system HR~6819 \citep[][but see also follow-up discussions: \citealt{bodensteiner2020,safarzadeh2020,elbadry2020}]{rivinius2020}. X-ray binaries are nearly all located within the mass transfer region and above the limit defined by Eq.~\ref{eq:ll}, except 4U 1543-475 which is known to have a low inclination $i\approx 20^\mathrm{o}$. 

For comparison, the latest selection of giant stars having large $\Delta V_r$ and optically invisible companions from the LAMOST survey \citep{zheng2019} are also shown. All, including GS4, lie in the grey area, indicating that the constraint on the radial velocity is not sufficient to reveal quiescent black holes given their relatively short orbital periods. To match the $M_{\rm bh} \geq 3\rm M_\sun$ criterion, these systems would need to have an upper limit on their inclination of $i \lesssim 65^\mathrm{o}$ for the source closest to the line \citep[source \#8 in][]{zheng2019} and between $i \lesssim 40^\mathrm{o}$ and $i \lesssim 20^\mathrm{o}$ for all the other candidates, if we assume the conservative $M_\star = 3\rm\,M_\sun$. However, at such low inclination we do not expect to detect the strong ellipsoidal modulations that are seen in the ASAS-SN light curves \citep[see e.g.][]{zheng2019}. Therefore, it is likely that most of these systems are also stellar binaries.

Non-interacting systems clearly lie at longer orbital periods, as supported by the recent detection of one such system using the APOGEE radial velocity survey \citep[][and follow-up discussion: \citealt{vandenheuvel2020,thompson2020}]{thompson2019}. A simple, conservative criterion to select  candidate black hole systems for follow-up is, using Eq.~\ref{eq:ll}
\begin{equation}
\Delta V_r \geq 120 \left(\frac{\Delta t}{30\,{\rm d}}\right)^{-1/3}\,\mathrm{km\,s}^{-1}
 \end{equation}
with $\Delta t$ the elapsed time between the maximum and minimum velocity. This is similar to how  \citet{thompson2019} prioritized their candidate list for follow-up, leading to the detection of a strong candidate. GS4, with an initial $\Delta V_r=100\,\mathrm{km\,s}^{-1}$ measured from the survey  \citep{gu2019}, is at the limit of the criterion although this is possibly lowered by a long $\Delta t$ (the six LAMOST observations are probably spread in time over several years). Surveys like LAMOST with a long time base are well-suited to probe the region of the diagram with detached giant stellar companions. However, as GS4 highlights, the long $\Delta t$  induced by the sampling may also lead to wrongly identify stellar binaries with short orbital period as black hole candidates. A workaround avoiding these false positives could be to sample spectra on various timescales.

\section{Conclusion}
\label{sec:conclu}

The large radial velocity variations detected in LAMOST spectra have been used to highlight potential black hole candidates \citep{gu2019,zheng2019}. We tested this selection by investigating further the parameters of the one system, GS4, having a known X-ray counterpart and presented by \citet{gu2019} as a likely black hole or neutron star system with mass transfer from the giant to the compact object. The time resolved spectra that we obtained from the Liverpool Telescope along with the ASAS-SN V-band light curve of GS4, allowed us to fully constrain the orbital parameters of this eclipsing binary, thereby excluding the possibility of a massive and compact companion. If our \textit{Chandra} observation did confirm the existence of an X-ray counterpart for this system, its X-ray spectrum and luminosity are fully consistent with a chromospherically active giant star, allowing us to classify GS4 as a RS CVn.

Given the radial velocity variations sampled by LAMOST $\Delta V_r \sim 100\rm\,km s^{-1}$ and the very high inclination of GS4 constrained from its eclipsing behavior, we argue that its short orbital period $P_{\rm orb} \approx 5.2\rm\,d$ was sufficient to exclude the black hole companion hypothesis for this system. The same reasoning likely applies to all other LAMOST candidates, except for one \citep[source \#8 in][]{zheng2019}, since they all have similar radial velocity variations, relatively short orbital periods and rather high inclinations implied from the orbital modulations seen in their ASAS-SN light curve \citep{zheng2019}. Therefore, the orbital period or the timescale of the radial velocity variations appear as key to check the viability of these radial velocity selected black hole candidates.

\begin{acknowledgements}
      The scientific results reported in this article are based on observations made with the Liverpool Telescope operated on the island of La Palma by Liverpool John Moores University in the Spanish Observatorio del Roque de los Muchachos of the Instituto de Astrofisica de Canarias with financial support from the UK Science and Technology Facilities Council, and on observations made by the \textit{Chandra} X-ray Observatory. MC and GD acknowledge financial support from the Centre National d’Etudes Spatiales (CNES). JC acknowledges support by the Spanish MINECO under grant AYA2017-83216-P.
\end{acknowledgements}

\bibliographystyle{aa} 
\bibliography{Clavel_GS4_accepted}


\end{document}